\documentclass[fleqn,11pt,a4paper]{article}
\usepackage[dvips]{graphicx}

\newcommand{\lsim}{\rlap{\raise 2pt \hbox{$<$}}{\lower 2pt \hbox{$\sim$}}}
\newcommand{\gsim}{\rlap{\raise 2pt \hbox{$>$}}{\lower 2pt \hbox{$\sim$}}}

\newcommand{\etal}{{\it et al.}}
\newcommand{\ie}{{\it i.e.\ }}
\renewcommand{\d}{\mbox{\rm d}}
\newcommand{\eg}{{\it e.g.\ }}

\begin{document}
\thispagestyle{empty}   
\noindent
TSL/ISV-96-0135   \hfill ISSN 0284-2769\\
March 1996       \hfill                \\
\\
\vspace*{5mm}
\begin{center}
  \begin{LARGE}
  \begin{bf}
High Energy Neutrino Production\\
by Cosmic Ray Interactions in the Sun\\
  \end{bf}
  \end{LARGE}
  \vspace{5mm}
  \begin{Large}
G.~Ingelman$^{a,b,}$\footnote{ingelman@tsl.uu.se} and 
M. Thunman$^{a,}$\footnote{thunman@tsl.uu.se} \\
  \end{Large}
  \vspace{3mm}
$^a$ Dept. of Radiation Sciences, Uppsala University,
Box 535, S-751 21 Uppsala, Sweden\\
$^b$ Deutsches Elektronen-Synchrotron DESY,
Notkestrasse 85, D-22603 Hamburg, Germany\\
  \vspace{5mm}
\end{center}
\begin{quotation}
\noindent
{\bf Abstract:}
The flux of neutrinos originating from cosmic ray interactions with
matter in  the Sun has been calculated based on Monte Carlo models for
high energy  particle interactions. The resulting flux at the Earth
(within the Sun's  solid angle) is higher than the corresponding one
from cosmic ray interactions  with the Earth atmosphere.  The smallness
of the absolute rate, however,  precludes it as a practical `standard
candle' for neutrino telescopes and  limits neutrino oscillation
searches. On the other hand, it facilitates dark matter searches based
on neutrinos from neutralino annihilation in the Sun. 
\end{quotation}

\section{Introduction}

High energy cosmic ray particles, mainly nucleons, interacts with
matter to produce secondary particles that are mainly mesons and some
baryons. This applies in particular to cosmic rays entering the
atmosphere of the Sun. The produced particles propagate through the Sun
until they either decay or make secondary interactions producing new
particles that contribute further to develop a cascade. Decays of
particles in such cascades will  produce neutrinos and other leptons,
\eg  muons that in turn decay into  neutrinos.  This scenario is
similar to the cascades induced by cosmic rays  in the Earth's
atmosphere, which we have studied extensively \cite{GIT}.  However, the
solar atmosphere is less dense at the typical interaction heights  and,
therefore, a larger fraction of the mesons will decay instead of 
interacting. This leads to relatively more neutrinos produced in the
Sun as compared to the Earth. 

It has been proposed \cite{Moskalenko} that the Sun might be used as a 
`standard candle' for neutrino telescopes, which is only possible if
the flux is significantly higher than the Earth's atmospheric flux.  A
large such neutrino flux from the Sun would, on the other hand, be a 
severe background for searches of neutrinos from neutralino
annihilation  in the Sun \cite{Kamionkowski}.  The hypothetical
neutralinos appear in theories based on  supersymmetry (SUSY)
\cite{SUSY} and are of fundamental interest in particle  physics as
well as in cosmology since they could contribute to the dark matter in
the Universe.  

In this paper we study the production of muon and electron neutrinos
in  cosmic ray interactions in the Sun, as well as their propagation
through the Sun and to the Earth where they could be detected in
neutrino telescopes,  such as {\sc Amanda} \cite{Amanda}, {\sc Baikal}
\cite{Baikal}, {\sc Dumand}  \cite{Dumand} and {\sc Nestor}
\cite{Nestor}. The cascade interactions in the  Sun are treated in
detail using Monte Carlo methods to simulate the high  energy particle
interactions. In particular, the Lund model \cite{Lund} and Monte
Carlo programs \cite{Pythia} are invoked.  The resulting neutrino
fluxes at the Earth are compared with the fluxes from  cosmic ray
interactions in the Earth's atmosphere \cite{GIT}.  This solar neutrino
flux is discussed in terms of the above `standard candle'  idea and
neutralino search. We also investigate the possibility of neutrino 
oscillations taking place between the source at the Sun and the
detector at the Earth. 

\section{Models and calculational techniques}

\subsection{Cosmic ray spectrum}
The flux of primary cosmic ray particles is conventionally parametrised
as \cite{GIT,Gaisser90,Lipari93}
\begin{equation}
\label{eq:initflux}
\phi_{N}(E)\left[\frac{\mbox{nucleons}}
{\mbox{cm$^{2}$\,s\,sr\,GeV/$A$}} \right] = \left\{ \begin{array}{ll}
    1.7\,E^{-2.7} &  E<5\cdot10^6\,\mbox{GeV} \\
      & \\
      174\,E^{-3} & 
      E>5\cdot10^6\,\mbox{GeV}
      \end{array}
      \right.
\label{eq:primary}
\end{equation}
The normalisation is here derived \cite{Pal92} from the directly 
measured primary spectrum using balloon-borne emulsion chambers in
JACEE \cite{JACEE}. It agrees (within some 10\%) with more indirectly
derived spectra based on measured atmospheric muon fluxes \cite{MACRO},
and is also compatible with the data discussed in ref.~\cite{Honda}.
For the energies of our interest ($E \gsim 100$\,GeV), we take the
cosmic ray flux to be isotropic, since the anisotropy is $\lsim 5\%$
\cite{Gregory82}.

The cosmic ray composition is dominated by protons with only a smaller
component of nuclei \cite{Pal92,Honda}. Likewise, the outer parts of
the solar atmosphere consist mainly of hydrogen with only a small
fraction of helium. Therefore, the interactions producing the secondary
particle fluxes are dominantly proton-proton collisions. The small
contribution of nuclear collisions can also be treated as
nucleon-nucleon interactions, since the nuclear binding energies are
negligible and other nuclear effects have little influence on the high
energy secondary particles which we are interested in.

\subsection{Solar matter distribution}

The interaction of cosmic ray particles in the Sun can be treated
analogously to our calculation of cosmic ray interactions in the
atmosphere of the Earth \cite{GIT}. However, the atmosphere of the Sun
is less dense and particles may therefore propagate deeper in to the
Sun such that a more elaborate matter density profile is needed.
Although there is no well defined solar surface, it is convenient to
consider different regions in relation to the solar radius  
$R_{\odot}=6.96\times 10^5 \: km$. One may then apply an exponential
density profile 
\begin{equation}\label{density}
\rho(h)=\rho_0\, e^{-h/h_0},
\end{equation}
where $h>0$ and $h<0$ corresponds to locations above and below
$R_{\odot}$, respectively. The parameters $\rho_0$ and $h_0$, given in
Table~\ref{tab:sun}, were obtained by fitting the data of
ref.~\cite{sunout} for the atmosphere and of ref.~\cite{sunin} for the
interior of the Sun. For muons which can reach deep into the Sun
($\sim 20\,000\,km$ vertical depth), a third region is introduced to
get an adequate description of the density  profile.

\begin{table}
\begin{center}
\begin{tabular}{|l|rr|}
\hline
height $h\: [km]$ & $\rho_0\: [g/cm^3]$ & $h_0\: [km]$ \\
\hline
$h>0$             & $3.68\cdot 10^{-7}$ & $115$   \\
$-2\,000<h<0$     & $3.68\cdot 10^{-7}$ & $622$   \\
$h<-2\,000$       & $45.3\cdot 10^{-7}$ & $2835$  \\  \hline
\end{tabular}
\end{center}
\caption{\em Parameters for the solar density profile in
Eq.\,(\ref{density}).}  \label{tab:sun}
\end{table}

To calculate the flux of particles from the Sun that reach the Earth
one  must also consider where in the Sun the primary cosmic ray
particle interacts. The essential point is whether the secondary
particles and final neutrinos  only have to pass through the solar
atmosphere or through the higher density interior before reaching the
Earth. This corresponds to an impact  parameter $b$ that varies
between $R_{\odot}$ (peripheral hit) and zero  (central hit) and a
correspondingly varying effective density function along the particle
trajectories. An exact treatment of this in our Monte Carlo  simulation
becomes quite complex. We have therefore simplified the situation  by
making the simulations for fixed impact parameter values  ($b=0$,
$b=2R_{\odot}/3$ (geometric average) and $b=R_{\odot}$)  and then
interpolated between them for the integrated results  taking neutrino
attenuation into account. 

\subsection{Basic particle interactions}
The interaction of the cosmic ray particles with the solar material is
treated as basic proton-proton collisions at high cms energy. The 
production of secondary particles and their decay into neutrinos can
then  be simulated in great detail using the Lund Monte Carlo programs 
{\sc Pythia} and {\sc Jetset} \cite{Pythia}. The neutrino flux arise
from  the decay of ordinary mesons, mainly $\pi$ and $K$, and from
decay of muons.

The production of ordinary hadrons (not containing heavy quarks) is
dominantly through minimum bias hadron-hadron collisions. The strong
interaction mechanism is here of a soft non-perturbative nature that
cannot be calculated based on proper theory, but must be modelled. In
the successful Lund model \cite{Lund} hadron production arise through
the fragmentation of colour string fields between  partons scattered in
semi-soft QCD interactions \cite{Pythia}. The essentially
one-dimensional colour field arising  between separated colour charges
is described by a one-dimensional  flux tube whose dynamics is taken as
that of a massless relativistic string.  Quark-antiquark pairs are
produced from the energy in the field through  a quantum mechanical
tunneling process. The string is thereby broken into smaller pieces
with these new colour charges as endpoints and, as the process is
iterated, hadrons are formed. These obtain limited momenta transverse
to the string (given by a Gaussian of a few hundred MeV width) but
their longitudinal momentum may be large as  it is given by a
probability function in the fraction of the available  energy-momentum
in the string system taken by the hadron. All mesons and  baryons in
the basic multiplets may be produced and the subsequent decays are
fully included.  The iterative and stochastic nature of the process is
the basis for the implementation of the model in the {\sc Jetset}
program \cite{Pythia}.

A non-negligible contribution to the inclusive cross section is given
by diffractive interactions.  These are also modeled in {\sc Pythia}
\cite{Pythia} using cross sections from a well functioning Regge-based
approach and  simulating the diffractively produced final state using
an adaptation of  the Lund string model. These diffractive events are
included in our  simulations and contribute rather less than 10\% to
the final results.  

In our study of the neutrino flux from cosmic ray interaction in the
Earth's atmosphere \cite{GIT}, an important point was the production of
charmed  particles. Their prompt decays gives a non-negligible
contribution to the  neutrino flux at very high energies ($E_{\nu}\gsim
10^{6}\,GeV$).  This is due to the increase in the $\pi$ and $K$ decay
lengths with increasing  energy, such that the probability for
interacting before they decay increase.  The lower density in the solar
atmosphere imply that their interaction length  is larger, such that
they still dominantly decay rather than being lost through
interactions. Therefore, so-called prompt neutrinos from charm decay 
are not as important as for the lepton fluxes in the Earth's
atmosphere. 

\subsection{Cascade evolution}
To describe the evolution of a cascade in the solar medium we use the
same  formalism as for interactions in the Earth's atmosphere and
therefore refer to our earlier study in ref.~\cite{GIT} for more
details. 

The flux of  nucleons in the solar atmosphere develops according to the
cascade equation 
\begin{equation}\label{N-flux}
\frac{\d\phi_N}{\d X}= -\frac{\phi_N}{\lambda_N} + S(N\,A\to N\,Y),
\end{equation}
where $X$ is the depth in the atmosphere and 
\begin{equation} \label{eq:intlength} \lambda_{N}(E) = { \rho(h) \over
\sum_A \sigma_{NA}(E) \, n_A(h) }, 
\end{equation}
is the nucleon interaction length in terms of the number density,
$n_A(h)$,  of nuclei $A$ at height $h$ and the nucleon-nucleus
inelastic cross section,  $\sigma_{NA}(E)$.   The first term in
Eq.\,(\ref{N-flux}) is the loss of nucleons due to interactions and
$S(N\,A\to N\,Y)$ is the regeneration due to interactions of primary 
nucleons with higher energies. 

This transport equation is solved through a cascade simulation
algorithm as follows.  A cosmic ray proton is generated with energy
drawn from a flat distribution in log\,$E$, and a weight assigned to
reproduce the shape of the primary  spectrum Eq.\,(\ref{eq:initflux}). 
This primary proton is then propagated down through the solar medium
according to Eq.\,(\ref{N-flux}) without the regeneration term $S(NA\to
NY)$ resulting in  the solution 
\begin{equation}
\phi(h) = \phi_0\ e^{-X/\lambda_N},
\end{equation}
From this one can obtain the height of the primary interaction
by solving the equation
\begin{equation}
\label{eq:pheight}
-\ln\frac{\phi(h)}{\phi_0}=-\ln R = \int^h_{\infty}\,\d h\,\sigma\,n(h)
\end{equation}
for $h$, where $R\,\in]0,1[$ is a uniformly distributed random number. 

A proton-proton interaction is then generated in full detail with 
{\sc Pythia} \cite{Pythia} resulting in a complete final state of
particles. Secondary particles are followed through the atmosphere
where they decay  or interact producing cascades.  Secondary nucleons
give a flux that is rather small compared to the primary flux and could
therefore be neglected as a first approximation.  We do, however,
include the main part of this effect by taking into account  secondary
nucleons that have an energy of at least 30\% of the primary one.
Nucleons with a lower energy give a negligible contribution compared to
the primary flux due to its steep energy spectrum
Eq.\,(\ref{eq:primary}).  These leading nucleons emerging in the
interactions are therefore allowed  to generate a secondary interaction
at a height calculated according to
\begin{equation}
\label{eq:secheight}
-\ln\frac{\phi(h)}{\phi(H)}=-\ln R' = \int^h_{H}\,\d h\,\sigma\,n(h),
\end{equation}
where $H$ is the production height of the nucleon. The procedure is
iterated until the energy of the leading nucleon from an interaction
falls below 30\% of the primary cosmic ray proton energy.

\begin{figure}[tb]
\begin{center}
\includegraphics[width=15cm]{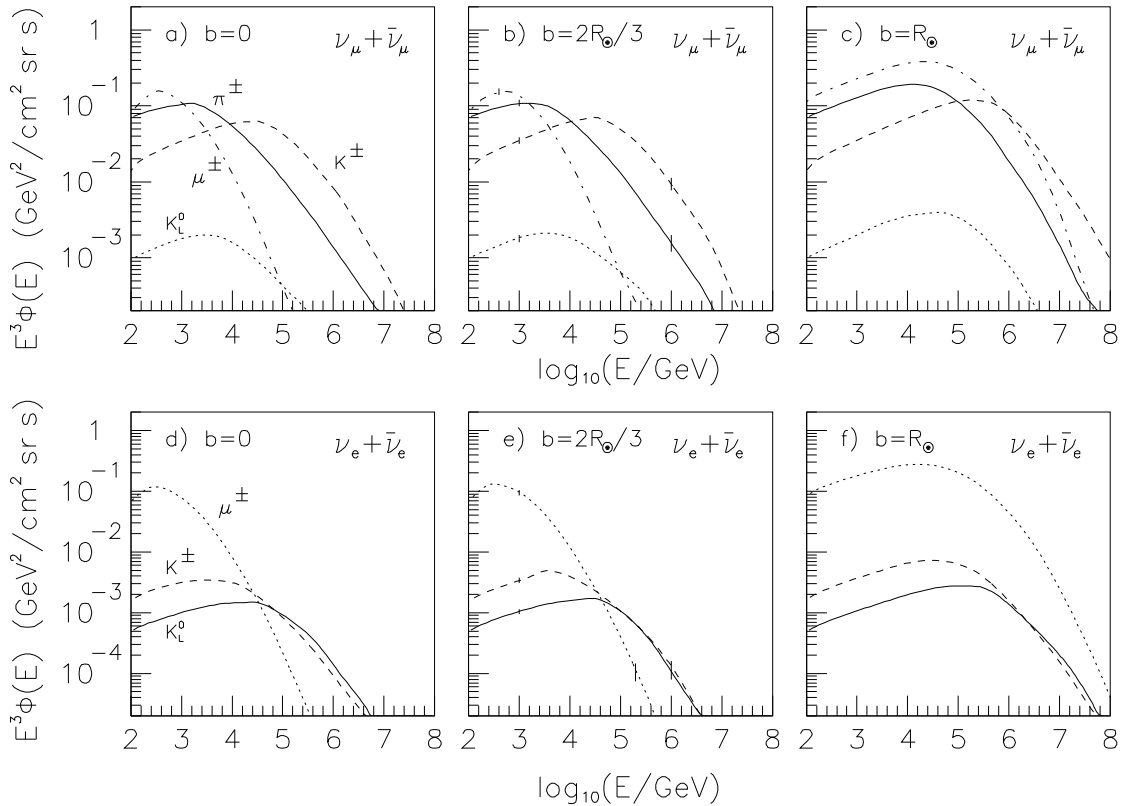}
\caption{\em 
The $E^3$-weighted flux of muon-neutrinos ($\nu_{\mu} +
\bar{\nu}_{\mu}$)  and electron-neutrinos ($\nu_e + \bar{\nu}_e$) from
decays of the specified  particles ($\pi , K, \mu$) produced in cascade
interactions in the Sun at different impact parameters b. The error
bars in (b,e) indicate the statistical precision (bin size 0.1 in
$log_{10}\,E$) of the Monte Carlo simulation.
Note, the neutrino flux attenuation from propagation through the Sun
(section 2.6) is not included.}
\label{fig:influx}
\end{center}
\end{figure}

The secondary mesons are propagated down into the Sun until they either
decay or interact, which is decided by comparing simulated interaction
and decay lengths. The interaction length is calculated analogous to
Eq.\,(\ref{eq:secheight}), while the decay length is given by 
\begin{equation}
L_{dec.} = -c\,\beta\,\gamma\,\tau\,\ln\,R\,.
\end{equation}
Particle decays are fully simulated with daughter particle momenta. In
case of interactions, the interacting particle is  regenerated in the
same direction but with degraded energy, chosen according to the
appropriate leading particle spectrum. Considering only the most
energetic `leading' particles in secondary  interactions is justified
because they give the dominant contribution to the high energy
neutrino flux. Other particles with lower energy are much fewer than
the particles of the same type and energy produced in  primary
interactions. Moreover, secondary interactions takes place deeper  in
the atmosphere, where the density is higher, giving a higher
probability  that a produced meson interacts rather than decays. This
further decreases  the importance of particles from secondary
interactions.

The particle decay--interaction chain is then repeated until all
particles have decayed or their energy fallen below a minimum of
$100\,GeV$. The energy spectra for neutrinos are finally obtained by 
simply counting the number of generated neutrinos and applying the 
weight assigned to the primary proton. The contributions from decays of
the different mesons are shown in Fig.\,\ref{fig:influx} for the
different impact parameters $b$. The contribution from charmed and
heavier mesons is not included, but are unimportant as will be
discussed in section~\ref{sec:Results}.

\subsection{Muon propagation in the Sun}

Since muons do not feel the strong force their propagation through
matter is quite different from the hadrons. Muons interact
electromagnetically and rarely experience hard interactions with a
single large energy loss. Instead they typically loose small amounts of
energy in each collisions, which can be added such that the energy
loss can be treated as a continuous process and parameterized in the
form \cite{PDB}
\begin{equation}
\label{eq:energyloss}
\frac{\d E}{-\d h}=-\alpha\,\rho\ -\ \beta\,\rho\,E,
\end{equation}
with $\alpha=0.0025\,$GeV/(g/cm$^2$) and  $\beta=4.0\cdot
10^{-6}\,$(g/cm$^2)^{-1}$. This continuous energy loss makes
the above treatment of decay versus interaction based on decay and
interaction lengths unsuitable. Instead, we  apply a method where small
steps in energy loss ($\Delta E/E\simeq 10\%$) are taken, under which
the energy loss rate is approximately constant  (\ie $\d E/\d h\simeq
const$). The decay probability in such a step can then be calculated
analytically and Monte Carlo simulated. If a muon survives such a step
in energy the procedure is repeated until the muon decays or gets an
energy lower than our $100\,GeV$ cut off.

The resulting neutrino flux from muon decays are shown  in
Fig.\,\ref{fig:influx} for the three different impact parameters. For
muon production with impact parameter $b=R_{\odot}$ the effective
matter thickness is very small resulting in no significant energy loss,
such that the muons effectively propagate without interactions. 

\subsection{Neutrino flux attenuation in the Sun}

\begin{figure}
\begin{center}
\includegraphics[width=10cm]{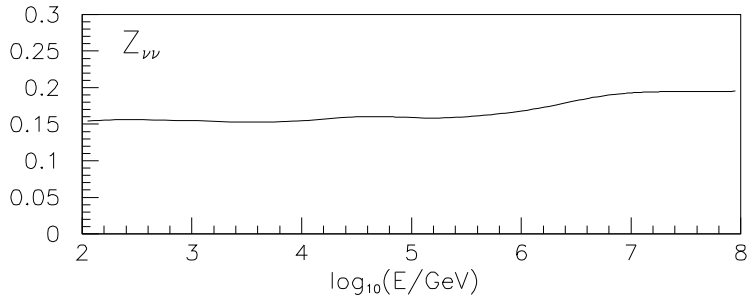}
\caption{\em 
Energy-dependence of the regeneration $Z$-moment, Eq.\,(\ref{Z-moment}) 
for neutrino-nucleon interactions.}
\label{fig:zmom}
\end{center}
\end{figure}

The neutrino flux will be attenuated through weak interactions with
the  nucleons of the solar medium. In charged current interactions the
neutrino  is lost altogether and instead the corresponding charged
lepton emerges,  which looses significant energy or is absorbed since
this process mainly  occurs in the dense interior of the Sun. In a
neutral current interaction the neutrino emerges with a reduced energy. 
We treat this using the analytic technique developed for atmospheric
cascades of hadrons \cite{GIT} and express the flux by  
\begin{equation}
\label{eq:cn}
\frac{\d \phi_{\nu}}{\d X} = - \frac{\phi_{\nu}} {\lambda_{\nu}} + 
S(\nu A\to \nu Y).
\end{equation}  
The first term, the absorption term, gives the loss due to both charged
and neutral current interactions. The second gives the
regeneration due to neutral current interactions of neutrinos of higher
energies. The latter term is given by
\begin{equation}
\label{eq:snn} 
S(\nu A\to \nu Y) = \int_{E}^{\infty} \d E' \, \frac{\phi_{\nu}(E')}
{\lambda_{\nu}(E')}  \, \frac{\d n_{\nu A\to \nu Y}(E',E)}{\d E} . 
\end{equation}
and can be rewritten as
\begin{equation}\label{Z-moment}
S(\nu A\to \nu Y) = \frac{\phi_{\nu}(E)}{\lambda_{\nu}(E)}
\int_{E}^{\infty} \d E' \, \frac{\phi_{\nu}(E')}{\phi_{\nu}(E)}\ 
\frac{\lambda_{\nu}(E)}{\lambda_{\nu}(E')}\  \frac{\d n_{\nu A\to \nu
Y}(E',E)} {\d E} = \frac{\phi_{\nu}(E)}{\lambda_{\nu}(E)} \, Z
\end{equation}
where the integral defines the regeneration $Z$-moment \cite{GIT}. To
use this formula it is normally assumed that the dependence on the depth $X$
cancels in the ratio $\phi_{\nu}(E')/\phi_{\nu}(E)$, which it does
if the interaction length is only weakly energy dependent. 
Although this is not the case here, this formula is still applicable since 
the $\d n/\d E$ spectra are peaked at low energy loss (for high energies).
Furthermore, one can use the same $Z$-moment for all neutrinos  ($\nu_e,
\bar{\nu}_e,\nu_{\mu}, \bar{\nu}_{\mu}$) since their spectral form 
before attenuation in the Sun are approximately the same and their 
cross sections are approximately the same. The $Z$-moment is calculated
with {\sc Pythia} \cite{Pythia} using a standard technique \cite{GIT}
and it is shown in Fig.\,\ref{fig:zmom}. Now, Eq.\,(\ref{eq:cn}) can be
solved to give the attenuated neutrino flux
\begin{equation}
\phi_{\nu}(r) = \phi_{\nu,0}\,\exp\, \{ -(1-Z)X(r)/\lambda_{\nu}(E_{\nu})
\, \},     
\end{equation}
where $X(r)$ is the amount of material being traversed.

The above treatment is of course a simplification. For example, the
flux, $\phi(E)$ is a function of the impact parameter $b$. Here, a fit
to the flux integrated over the Sun is used. Compared to the two most
rough simplifications that can be done, neglecting secondary neutrinos
and neglecting energy loss in neutral current interactions, our
simplifications are reasonable. By comparing the $Z$-moments, which are
zero if secondary neutrinos are neglected and $ \sigma^{NC}/
\sigma^{NC+CC}\,\sim\,2/7$ if energy loss is neglected in neutral
current interactions, we can see that the above treatment is a
significant improvement.

\begin{figure}[bth]
\begin{center}
\includegraphics[width=8cm]{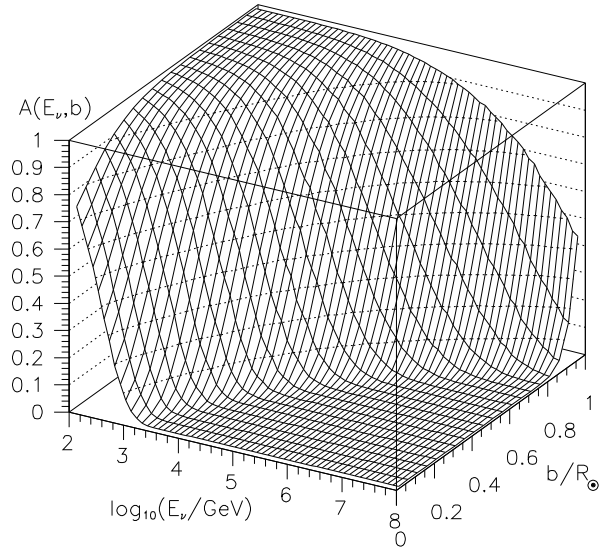}
\caption{\em The attenuation factor, Eq.\,(\ref{eq:supnu}), for
neutrinos  passing through the Sun as a function of neutrino energy and
impact parameter $b$.}
\label{fig:supnu}
\end{center}
\end{figure}

In principle, each neutrino should be followed through the Sun with the
attenuation folded in to get the final attenuated spectrum. However,
since the neutrino production only occur in a tiny outer fraction of
the Sun one may use a simpler factorised approximation. The previously
obtained fluxes can thus be multiplied by an overall attenuation
function for the Sun. This attenuation function can then be calculated
in terms of the impact parameter 
\begin{equation}
A(E_{\nu},b)=\exp\left\{-\sigma(E_{\nu})\,(1-Z)\,X(b)/m_N\right\}.
\label{eq:supnu}
\end{equation}
where $X(b)$ is the effective thickness of the Sun and $m_N$ is the
nucleon  mass. A numerical evaluation of this function is shown in
Fig.\,\ref{fig:supnu}.

\section{Resulting neutrino fluxes}
\label{sec:Results}

\begin{figure}[tb]
\begin{center} 
\includegraphics[width=13cm]{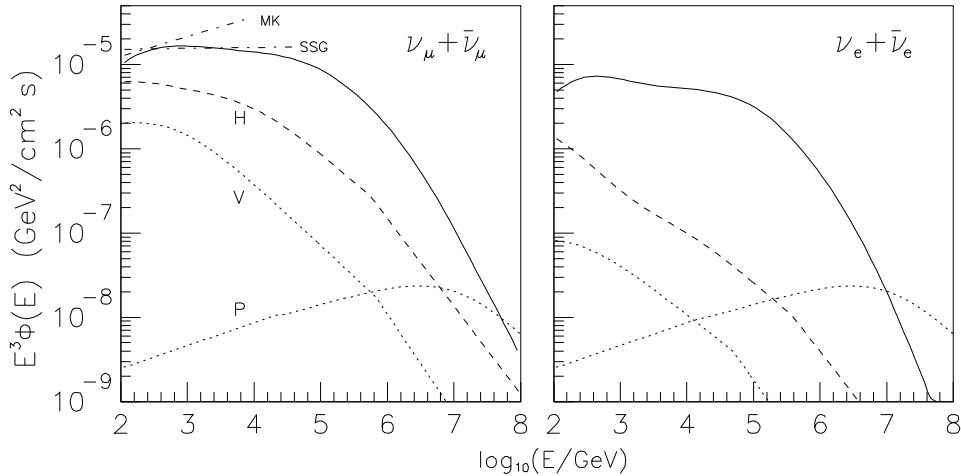}
\caption{\em  Cosmic ray induced $E^3$-weighted neutrino fluxes at the
Earth integrated over  the solid angle of the Sun. The fluxes from the
Sun obtained in this study  (solid lines) are compared with the earlier
calculation SSG \protect\cite{SSG}  and the one MK derived from
\protect\cite{Moskalenko}, as well as those from  the Earth's
atmosphere as calculated for   the vertical flux (curve V)
\protect\cite{GIT},  the horizontal flux (curve H)
\protect\cite{Volkova79}, and the prompt charm-induced flux (curve P)
\protect\cite{GIT}.}
\label{fig:earthflux} 
\end{center} 
\end{figure}

The unattenuated neutrino flux from the decay of different particles 
produced in cosmic ray collisions in the Sun are shown in 
Fig.\,\ref{fig:influx}. As can be seen from the results for different 
impact parameter values, secondary interactions of mesons and energy
loss for the muons are important and give substantially less high
energy neutrinos at more central interactions. By interpolating over
different impact parameters and folding with the attenuation factor,
Eq.\,(\ref{eq:supnu}), the integral over the solar disc is
carried out resulting in the total neutrino fluxes at the Earth shown
(by solid curves) in Fig.\,\ref{fig:earthflux}. The fluxes are compared
with the horizontal \cite{Volkova79}, the vertical \cite{GIT} and the
prompt \cite{GIT} fluxes from cosmic ray interactions in the Earth's
atmosphere. 

As demonstrated in Fig.\,\ref{fig:earthflux}, the fluxes from the Sun
are significantly higher than those from the Earth's atmosphere. The
lower density of the solar atmosphere gives a lower probability for
secondary interactions and hence favours decays into high energy
neutrinos. For the muon neutrinos, the solar flux is thus about one
(two) orders of magnitude larger than the horizontal (vertical)
atmospheric flux in the energy range $10^4$--$10^7\,GeV$. In the case
of electron neutrinos, the solar flux very much higher at  all
energies. This is due to the contribution from muon decays (see
Fig.\,\ref{fig:influx}), which is only important for the very lowest 
energies in the atmospheric fluxes. For both muon and electron
neutrinos, the slope at high energies ($\gsim 10^6\,GeV$) is steeper
for the solar fluxes compared to atmospheric ones. This is because of
neutrino attenuation, Eq.\,(\ref{eq:supnu}) and Fig.\,\ref{fig:supnu},
is here getting noticeable in the Sun. 

As mentioned earlier the prompt neutrino contribution, \ie from hadrons
with charm and heavier quarks, has not been explicitly studied. The
prompt atmospheric fluxes \cite{GIT} are plotted in
Fig.\,\ref{fig:earthflux} (curve P) for comparison. Due to the short
charmed particle life-time, these fluxes only depend on the production
rate up to $\sim 10^7\,GeV$. The production in the Sun would therefore
be the same, but the attenuation in the Sun results in a prompt
neutrino flux that is lower than the atmospheric one.

Our solar muon neutrino flux is, in Fig.\,\ref{fig:earthflux}, compared
with some earlier calculations. The results of Seckel, Stanev and
Gaisser \cite{SSG} are in agreement both regarding normalisation and
shape, although  they do not extend to as large energies as our
calculation. The result of Moskalenko and Karakula \cite{Moskalenko}
shows a significant difference in shape, related to their neglect of 
secondary interactions. By comparing our Fig.\,\ref{fig:influx}ab with
Fig.\,\ref{fig:influx}c it is obvious that secondary interactions and
energy loss for muons cannot be neglected, even at relatively low
energies. In fact by using the fluxes in Fig.\,\ref{fig:influx}c (where
secondary interactions are unimportant) for the whole Sun, the result
of Moskalenko and Karakula is roughly reproduced.

One can represent the neutrino fluxes by the simple parameterisation  
\begin{equation} 
\label{eq:fiteq} 
\phi(E)=\left\{\begin{array}{ll}
N_{0}\,E^{-\gamma-1}/(1+A\,E)\,, & E<E_{0} , \\ & \\
N_{0}'\,E^{-\gamma'-1}/(1+A\,E) \,, & E>E_{0} . \\ \end{array} \right.
\end{equation}
Although the form is the same as in Eq.\,(19) of \cite{GIT} for the  
atmospheric fluxes, the physical interpretation is not as simple here 
due to the integration over impact parameter and the inclusion of the 
attenuation factor. Still, this form gives a good fit to the total
attenuated  fluxes in Fig.\,\protect\ref{fig:earthflux} resulting in
the  parameter values in Table\,\ref{tab:fit}. ($N_0'$ is not fitted
but given by the continuity condition at $E_0$.) 

\begin{table}[bht]
\begin{center}
\begin{tabular}{|l|llllll|}
\hline
  & \multicolumn{0}{c}{$N_{0}$}  & \multicolumn{0}{c}{$\gamma$} &
\multicolumn{0}{c}{$A$}          & \multicolumn{0}{c}{$E_{0}$} & 
\multicolumn{0}{c}{$\gamma'$}    & \multicolumn{0}{c|}{$N_{0}'$} \\
\hline
$\nu_{\mu}+\bar{\nu}_{\mu}$  & $1.3\cdot 10^{-5}$ & $1.98$ & 
$8.5\cdot10^{-6}$ & $3.0\cdot 10^{6}$ & $2.38$ & $5.1\cdot10^{-3}$ \\
$\nu_{e}+\bar{\nu}_{e}$ & $7.4\cdot 10^{-6}$  &  $2.03$ &
$8.5\cdot10^{-6}$  &  $1.2\cdot 10^{6}$  &  $2.33$  & 
$5.0\cdot10^{-4}$ \\ \hline
\end{tabular}
\end{center}
\caption{\em Values of parameters in Eq.\,(\protect\ref{eq:fiteq})
obtained  from fits to the attenuated neutrino fluxes given by the
solid lines in Fig.\,\protect\ref{fig:earthflux}.}
\label{tab:fit}
\end{table}

\begin{figure}
\begin{center}
\includegraphics[width=8cm]{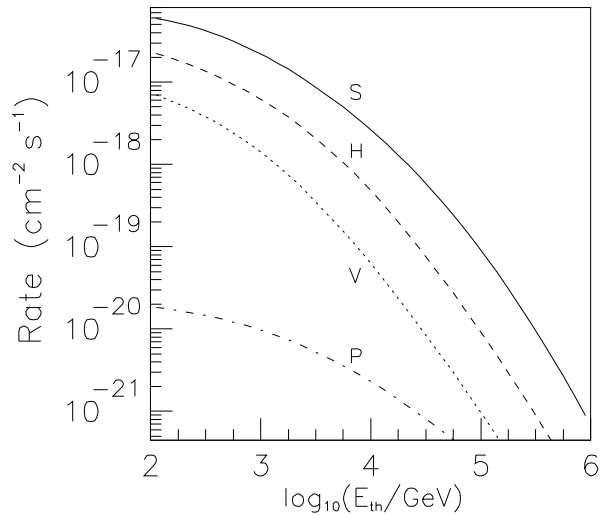}
\caption{\em The integrated rate of muon neutrino events in a neutrino 
telescope, as given by Eq.\,(\protect \ref{eq:intweight}).  The event
rate for neutrinos from the Sun (S) as compared to the ones based  on
the horizontal (H), vertical (V) and prompt (P) Earth atmospheric
fluxes.} \label{fig:rateflux}
\end{center}
\end{figure}

Neutrino telescopes measure neutrino fluxes indirectly through  the
\c{C}erenkov light emitted from the muons and electrons produced in 
charge current neutrino interactions.  The rate is, therefore, not
directly proportional to the flux, but rather  to the flux folded with
the probability that the neutrino undergoes a charge  current
interaction and that the produced charged lepton reaches the detector,
\ie proportional to its range. Thus, the rate in a water/ice
\c{C}erenkov telescope is given by the quantity
\begin{equation}
\label{eq:intweight}
R = \int^{\infty}_{E_{th}}\,\d E\,\phi(E)\,\sigma^{\nu\to\ell}(E)
\, \frac{\rho}{m_p}L(E) \, .
\end{equation}
Here, $\rho$ is the density of the medium ($\rho \approx\,1\,$g/cm$^3$
for water/ice), $m_p$ is the proton mass such that $\rho /m_p$ is the
target number density, and $L(E)$ is the range of the lepton. The
lepton energy is not the same as the neutrino energy, but for the
energies in this study it is a sufficiently good approximation. The
muon range is given by
\begin{equation}
L(E_{\mu}) = \frac{1}{\beta\,\rho}\,\ln\left( \frac{E_{\mu}\,+
\,\alpha/\beta} {\alpha/\beta} \right)
\end{equation}
with $\alpha,\: \beta$ from the muon energy loss formula 
Eq.\,(\ref{eq:energyloss}). The results of a numerical evaluation of the
rate, Eq.\,(\ref{eq:intweight}), are shown in Fig.\,\ref{fig:rateflux}. 

The rate that this flux would induce in neutrino telescopes under
construction is very low. For example, the version of {\sc Amanda} 
currently being deployed will have an effective area of about $3\cdot
10^8\,cm^2$ and a neutrino energy threshold of about $100\,GeV$. This would
give one event per two years of running. 

\section{Neutrino oscillations}

If neutrinos are not massless there may be oscillations between the
different weak eigenstates $\nu_e,\: \nu_{\mu},\: \nu_{\tau}$. This
phenomenon has been used to explain deficits of low energy solar
neutrinos and atmospheric neutrinos in the GeV energy range. A number of 
particle physics experiments at accelerators and nuclear reactors have
searched for oscillations. For a recent review of the field, see
\cite{Camilleri94}. 

If neutrino oscillations indeed occur, they will affect also the high
energy  neutrino fluxes from the Sun. The situation is, however, very
different from  the case of low energy solar neutrinos, where the
Mikheyev-Smirnov-Wolfenstein (MSW) effect \cite{MSW} is the dominating
source of oscillations \cite{Hata94}. The origin of the MSW-effect is
the different `index of refraction' for muon  and electron neutrinos in
matter and it only occurs when certain conditions for neutrino energy
and nuclear number density are satisfied. These conditions are not
fulfilled for the high energy neutrinos considered here and, since the
fluxes are strongly attenuated when passing through the Sun
(Fig.\,\ref{fig:supnu}), the effect can be  neglected.  The
oscillations could instead be of importance due to the large distance 
from the Sun as source to the detector at the Earth. 

In our analysis of this, we assume that there are two massive
neutrino  states that mixes. The probabilities that a neutrino emitted
as flavour $\ell$ is of flavour $\ell$ or $\ell '$ at detection are
given by 
\begin{eqnarray}
P_{\nu_{\ell}\nu_{\ell}}(E_{\nu},x) = 1 - \sin^2(2\,\!\theta)\
\sin^2\left(\frac {\delta\,\!m^2 x}{4\,E_{\nu}}\right) & & \\
P_{\nu_{\ell}\nu_{\ell'}}(E_{\nu},x) = \sin^2(2\,\!\theta)\
\sin^2\left(\frac{\delta\,\!m^2 x}{4\,E_{\nu}}\right),& & 
\end{eqnarray}
where $\theta$ is the neutrino mixing angle,
$\delta\,\!m^2\,=\,|m_1^2-m_2^2|$  is the difference of the squared
neutrino masses and $x$ is the traversed  distance, \ie $x=D$ for the
Sun-Earth distance.  

\begin{figure}
\begin{center}
\includegraphics[width=15.8cm]{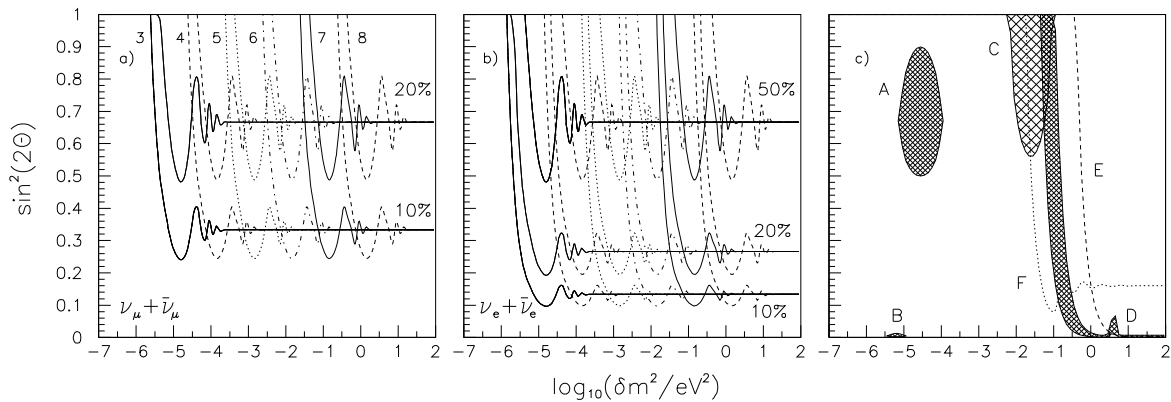}
\caption{\em  Iso-lines for the indicated percentage (a) increase of
the muon neutrino flux and (b) decrease of the electron neutrino flux
due to $\nu_{\mu}$--$\nu_e$ oscillations as functions of the neutrino
mixing angle $\theta$ and mass  difference
$\delta\,\!m^2\,=\,|m_1^2-m_2^2|$. Results, see Eq.\,(\ref{eq:osc}), 
averages over energy bins in $\log \left\{ E_{\nu} \right\}$ centered
at  $E_{\nu}=10^n \: GeV$ with the indicated $n=3,4...8$.  (c) Limits
and not excluded regions based on different experiments, see text.}
\label{fig:oscillation}
\end{center}
\end{figure}

Fig.\,\ref{fig:oscillation} illustrates the effect of the neutrino
oscillation as a function of these two basic oscillation parameters by
showing  iso-lines of the indicated per cent changes of the neutrino
fluxes, \ie in the ratio  
\begin{equation}\label{eq:osc}
\left\{ \Phi_{\nu_{\mu}}(E_{\nu}) P_{\nu_{\mu}\nu_{\mu}}(E_{\nu},D)
     +  \Phi_{\nu_{e}}(E_{\nu}) P_{\nu_{e}\nu_{\mu}}(E_{\nu},D) \right\}
     / \Phi_{\nu_{\mu}}(E_{\nu})
\end{equation}
for the muon neutrino and similarly for the electron neutrino. The
results are averages over energy bins in $log(E_{\nu})$
centered at $E_{\nu}=10^n\,GeV$ with $n=3,4...8$ as indicated. Given
the starting point with a larger muon neutrino flux than the electron
neutrino flux, see Fig.\,\ref{fig:earthflux}, we show the 10\% and 20\%
decrease of the muon neutrino flux and the 10\%, 20\% and 50\% increase
of the electron neutrino flux in Fig.\,\ref{fig:oscillation} (a) and
(b), respectively.

Two main features can here be observed. First, the lower limit of
$\delta\,\!m^2$ increases with increasing energy. Second, all curves
coincide at the highest $\delta\,\!m^2$, where the oscillation length
becomes much shorter than the Sun-Earth distance such that 
$sin^2(\delta\,m^2\ D/4\,E_{\nu})$ will be averaged out to $1/2$. 

Fig.\,\ref{fig:oscillation}c illustrates experimental limits and
acceptable  regions. Region $A$ and $B$ are, together with a very small
region around $sin^2 2\theta\,\sim\,1$ and $\delta m^2\sim
\,5\cdot10^{-8} eV^2$, the regions which could explain the solar
neutrino problem \cite{Hata94}. The region marked $C$ is allowed based
on observation of atmospheric neutrino fluxes \cite{Fukuda94}. $D$ is
the region not excluded by the LSND experiment \cite{LSND95}. The limit
$E$ is set by the Goesgen reactor experiment \cite{Zacek85} and $F$ by
the accelerator experiment LAMPF E734 \cite{Ahrens85}. Since these
allowed regions are not overlapping, one can interpret (neglecting
experimental uncertainties) that as either some result is incorrect or
there are more neutrino species involved in the oscillation.

Given the low absolute flux of high energy solar neutrinos, the event
rate  in currently planned neutrino telescopes will be too low for
revealing searches for, or studies of, neutrino oscillations. 

\section{Conclusions} 
\label{sec:Summary} 

We have calculated the high energy muon and electron neutrino fluxes 
arising from the interactions of cosmic ray particles with the solar
matter. Our resulting muon neutrino flux agrees with that obtained by
Seckel \etal\  \cite{SSG}, but our result extends a few orders of
magnitude higher in energy. The muon neutrino flux in
ref.~\cite{Moskalenko} is in disagreement with  both these results, due
to an oversimplified model where secondary interactions in the Sun are
not taken into account. 

These solar neutrino fluxes are one to two orders of magnitude larger
than  those from cosmic ray interactions in the Earth's atmosphere,
when integrated  over the solid angle of the Sun as seen from the
Earth. This opens a  possibility to use the solar neutrino flux as a
`standard candle' for  neutrino telescopes. However, one must here also
consider the angular spread introduced by the charged current
interaction producing the detectable muon relative to the incoming
neutrino direction.  This deflection is typically
$\sim10^{\circ}\sqrt{10\,GeV/E_{\mu}}$. In addition, the experimental
measurement of the muon direction also has a limited resolution. Taking
these two effects together, the solar disc will typically  cover less
than $\sim10\%$ of the solid angle that has to be integrated over. The
solar muon neutrino flux, which stays the same, would then have to be 
compared with a factor ten, or more, increased atmospheric flux such
that the two  would be of the same order of magnitude. Under these
conditions, one would  only have a factor two increase towards the Sun
and the use of  the Sun as `standard neutrino candle' does not look so
promising. One should here also be aware of the very low absolute rate
of events in  a neutrino telescope of the size now under
consideration.  For example, we estimate the rate of neutrino events
with $E_\nu >100\: GeV$ to be one per year in a detector covering
$6\cdot 10^4\: m^2$.  

We have also investigated the potential for observing neutrino
oscillations  during the passage from the source in the Sun and a
detector at the Earth. In principle, one could access an interesting
region in the parameter plane  of $sin^22\theta$ and $\delta m^2$, but
with the very low absolute rates it is beyond present neutrino
telescopes.

The positive consequence of these small solar neutrino fluxes is that
they will cause less of a background problem in attempts to detect
neutrinos from other sources. Of particular interest here is the search
for neutrinos from neutralino annihilation in the Sun, where the
predicted rate can be up to an order of magnitude larger depending on
the supersymmetric parameters \cite{Edsjo96}. A clear observation of
this phenomenon would both demonstrate supersymmetry, \ie physics
beyond the standard model in particle physics, and the presence of
non-baryonic dark matter in the Universe.

\end{document}